\documentclass[fleqn,twoside]{article}
\usepackage{latexsym,amsmath}
\begin{document}
\newcommand{\bbox}{\stackrel{-}{\Box}}
\newcommand{\bnabla}{\bar {\nabla}}
\begin{center}
INFLATIONARY DILATONIC de SITTER UNIVERSE   FROM  $ \cal {N}$ =  4    SUPER YANG - MILLS  THEORY PERTURBED BY SCALARS\\
\bigskip
John Quiroga\footnote{E-mail: jquiroga@tspu.edu.ru}\\
Department of Physics\\
Universidad Tecnol\'ogica de Pereira\\
Colombia\\
\bigskip
and\\
\bigskip Lab. for Fundamental Study,\\ Tomsk State Pedagogical
University, \\Tomsk
634041, Russia\\
\end{center}
\bigskip
\begin{center}
December 2002\\
\end{center}
\begin{abstract}
In this paper a quantum $\cal{N}$  = 4  super Yang-Mills theory
perturbed by dilaton-coupled scalars, is considered. The induced
effective action for such a theory is calculated on a
dilaton-gravitational background using the conformal anomaly found
via  AdS/CFT correspondence. Considering such an effective action
(using the large N method) as a quantum correction to the
classical gravity action with cosmological constant we study the
effect from dilaton to the scale factor (which corresponds to the
inflationary universe without dilaton ). It is shown that,
depending on the initial conditions for the dilaton, the dilaton
may slow down, or accelerate, the inflation process. At late
times, the dilaton is decaying exponentially. At the end of this
work, we consider the question how the perturbation of the
solution for the scale factor affects the stability of the
solution for the equations of motion and therefore the stability
of the Inflationary Universe, which could be eternal.
\end{abstract}
\newpage
It is known  that the Inflationary Universe (for a general review,
see \cite{KT})has been considered quite realistic as an element of
the very early  Universe evolution. However, evidences, recently
appeared, show that the present Universe is subject to accelerated
expansion  and thus may be entering an inflationary phase now.
Because of this, one may think that a reconsideration of quantum
cosmology and construction of  new (or modified) versions of the
theory of the very early Universe is possible.

In the present work, we consider one of such theories, which have
become very popular recently  in connection with AdS/CFT
correspondence, namely, quantum cosmology as induced from $\cal
{N}$ = 4 quantum super YM theory.  Using conformal anomaly on a
dilaton-gravitational background the anomaly induced effective
action was constructed in \cite{brevik99}and the consequences it
may lead to in the early Universe were discussed in
\cite{brevik99}. The present paper represents the generalization
of such model for quantum cosmology \cite{brevik99} where $N=4$
SYM is perturbed by quantum scalar fields interacting with
dilaton. Taking into account that on a purely gravitational
background such an effective action leads to the possibility of
inflation, we will show that the role of the dilaton is to
accelerate, or to slow down, the inflationary expansion, depending
on the choice of initial conditions for the dilaton.

 Let us start from the Lagrangian of local superconformally
 invariant  $\cal {N}$ = 4  super YM  theory in the background of
 $\cal{N}$ = 4  conformal supergravity. The corresponding vector multiplet is
 $(A_\mu, \psi_i, X_{ij})$. Supposing that super  YM  theory
 interacts with conformal supergravity in a  SU(1,1)  covariant
 way and keeping only kinetic terms, we get:
\begin{eqnarray}
L_{SYM}&=&-\frac{1}{4}(e^{-\phi}F_{\mu\nu}F^{\mu\nu}+\tilde{C}F^{\mu\nu}F_{\mu\nu}^*)- \nonumber  \\
       & &-\frac{1}{2}\bar{\psi}^i \gamma^\mu D_\mu \psi^i-
\frac{1}{4}X_{ij}(-D^2+\frac{1}{6}R)X^{ij}+...
\label{1}\end{eqnarray}

Let us remark that the scalar $\phi$ from the conformal
supergravity multiplet is written as $\tilde{C}+ie^{-\phi}$. It is
also to be noted that the first term in (\ref{1}) describes the
dilaton coupled electromagnetic field whose conformal anomaly has
been found in \cite{NO}, $\phi$ is complex scalar(dilaton).

Taking into account that in our theory we add scalars to the
action (\ref{1}), we must include into the action the term
describing it. Therefore for the $4d$ dilaton coupled scalars,
\begin{equation}
L=f(Re\phi)g^{\mu\nu}\partial_{\mu}X^k\partial_{\nu}X^k \,, \qquad
k=1,...,M \label{scalar}\end{equation}

Note  that $f$ is some function of real part of dilaton.

On a purely bosonic background with only non-zero gravitational
and dilaton fields, the conformal anomaly for ${\cal N}$ = 4 super
YM theory has been calculated in \cite{NO1} via AdS/CFT
correspondence \cite{MA} to be (adding conformal anomaly for
dilaton coupled scalar);
\begin{eqnarray}
T&=&b(F+\frac{2}{3}\Box R)+b'G+b''\Box R +C[\Box \phi^*\Box
\phi-2(R^{\mu\nu}-\frac{1}{3}g^{\mu\nu}
R)\nabla_{\mu}\phi^*\nabla_{\nu}\phi)]+\nonumber
\\&+&a_1\frac{[(\nabla f)(\nabla f)]^2}{f^4}+a_2 \Box
\left(\frac{(\nabla f)(\nabla f)}{f^2}\right).
\label{2}\end{eqnarray}

Here
\[ b=\frac{N^2-1}{(4\pi)^2}\frac{N_s+6N_f+12N_v}{120}+\frac{M}{120(4\pi)^2}=\frac{N^2-1}{4(4\pi)^2}+\frac{M}{120(4\pi)^2}, \]
\[b'=-\frac{N^2-1}{(4\pi)^2}\frac{N_s+11N_f+62N_v}{360}-\frac{M}{360(4\pi)^2}=-\frac{N^2-1}{4(4\pi)^2}-\frac{M}{360(4\pi)^2}, \]
\[ C=\frac{N^2-1}{(4\pi)^2}N_v=\frac{N^2-1}{(4\pi)^2}. \]
\[a_1=\frac{M}{32(4\pi)^2}\;\;,\qquad a_2=\frac{M}{24(4\pi)^2}\]

In the above expression for the anomaly we have taken into account
the fact that $N_s=6$, $N_f=2$, $N_v=1$ in ${\cal N}$ = 4 SU(N)
super YM theory; \(
F=R_{\mu\nu\alpha\beta}R^{\mu\nu\alpha\beta}-2R_{\mu\nu}R^{\mu\nu}+\frac{1}{3}R^2$
is the square of the Weyl tensor in four dimensions; $G$ is the
Gauss-Bonnet invariant. The prefactor $N^2-1$ appears because all
fields are in the adjoint representation. The conformal anomaly
for four-dimensional dilaton-coupled scalar has been found in
refs. \cite{SNSD}.

Note that both $N$ and $M$ maybe considered to be big parameters.
So, one can study large-$N$ or large-$M$ expansion.

Let us now find the anomaly induced effective action \cite{RFT}
(for review, see \cite{BOS}). We will write it in the
non-covariant, local form:
\begin{eqnarray}
W   &  =    &  b \int d^4 x \sqrt{-\bar{g}} \bar{F} \sigma+
   b'\int d^4 x \sqrt{-\bar{g}} [ \, \sigma [2\,{\bbox}^2 +\nonumber \\
    &  +    &     4 \bar{R}^{\mu \nu} {\bnabla}_\mu {\bnabla}_\nu
   -\frac{4}{3}\bar{R} \bbox  +\frac{2}{3}({\bnabla}^\mu \bar{R}){\bnabla}_\mu ]\sigma
     +(\bar{G}- \frac{2}{3} {\bbox} \bar{R})\sigma ]  -\nonumber \\
    & -  & \frac{1}{12}[b''+\frac{2}{3}(b+b')] \int d^4 x \sqrt{-\bar{g}}[ \,
\bar{R} -6 \bbox \sigma -6( {\bnabla}_\mu \sigma )({\bnabla}^\mu \sigma) ]^2 +\nonumber \\
    &  + & C \int d^4 x \sqrt{-\bar{g}}\, \sigma \phi^* ( {\bbox}^2 +2 {\bar{R}}^{\mu \nu}
{\bnabla}_\mu {\bnabla}_\nu-\frac{2}{3} \bar{R} \bbox
+\frac{1}{3}({\bnabla}^\mu \bar{R} ){\bnabla}_\mu)\phi + \nonumber
\\ &  + & \int d^4 x \sqrt{-\bar{g}}\, \{ a_1\frac{[(\nabla
 f)(\nabla f)]^2}{f^4}\sigma + a_2 \Box
\left(\frac{(\nabla f)(\nabla f)}{f^2}\right)\sigma +\nonumber
\\ & + &a_2
\frac{(\nabla f)(\nabla
f)}{f^2}\left[(\nabla\sigma)(\nabla\sigma)\right]\}.
\label{4}\end{eqnarray}

Note that in the conformal anomaly (\ref{2}) we used $ g_{\mu\nu}
= e^{2 \sigma} \bar{g}_{\mu \nu} $, and all quantities in
(\ref{4}) are calculated with the help of the overbar metric.

Since we know that the anomaly induced effective action is defined
with accuracy up to a conformally invariant functional, we may
limit ourselves to a conformally flat metric, i. e. $\bar{g}_{\mu
\nu}= {\eta}_{\mu \nu} $. In this case, the conformally invariant
functional on a purely gravitational background is zero, and  $ W
$ in Eq. (\ref{4}) gives the complete contribution to the one-loop
effective action.  In addition to this we will assume that only
the real part of the dilaton coupled to SYM theory is non-zero .

The anomaly induced effective action (\ref{4}) may now be
simplified significantly (due to the fact that $ {\bar{g}}_{\mu
\nu} = {\eta}_{\mu \nu} $):
\begin{eqnarray}
W   &  =    & \int d^4 x \{ \, 2 b' \sigma {\Box}^2 \sigma -3 ( b'' +\frac{2}{3}(b+b')) \times \nonumber \\
    &   & \times (\Box \sigma + \partial_\mu \sigma \partial ^\mu \sigma)^2
+ C \sigma \,\phi \, {\Box}^2 \phi + a_1\frac{[(\nabla
 f)(\nabla f)]^2}{f^4}\sigma  +\nonumber \\ &  & + a_2 \Box
\left(\frac{(\nabla f)(\nabla f)}{f^2}\right)\sigma +a_2
\frac{(\nabla f)(\nabla
f)}{f^2}\left[(\nabla\sigma)(\nabla\sigma)\right]\},
\label{5}\end{eqnarray}

where all derivatives are now flat ones.

Considering the case when the scale factor $ a(\eta) $ depends
only on conformal time: $\sigma(\eta)=\ln a(\eta) $, one has to
add the anomaly induced effective action to the classical
gravitational action:
\begin{equation}
S_{cl} =- \frac{1}{\kappa} \int d^4 x \sqrt{-g}\, (R+6\Lambda) =
-\frac{1}{\kappa}\int d^4 x e^{4\sigma}
(-6e^{-2\sigma}((\sigma')^2+(\sigma''))+6\Lambda),
\label{6}\end{equation}

where $ \kappa = 16 \pi G $.

Now, the equations of motion for the action $S_{total}=S_{cl}+W$,
assuming that $\sigma $ and $ \phi $ depend only on the conformal
time $\eta$, may be written in the following form (assuming the
simplest choice $f(Re\phi)=\phi$):
\begin{eqnarray}\label{eqmov}
\!\!\!\!\!\!\!\!&&\frac{a''''}{a} -   \frac{4
a'\,a'''}{a^2}-\frac{3 a''^2}{a^2}+\frac{a''\,a'^2}{a^3} \left( 6-
\frac{12 b'}{3b''+2 b}
\right) + \frac{12 b'\,a'^4}{(3 b''+2 b)a^4}- \nonumber \\
          &  & -\frac{6}{\kappa
(3b''+2b)} a\,a'' + \frac{12\Lambda}{\kappa
(3b''+2b)}a^4-\frac{C}{2(3b''+2b)}\phi\,\phi''''-\frac{a_1}{2(3b''+2b)}\frac{\phi'^4}{\phi^4}-
\nonumber \\ & &
-\frac{a_2}{2(3b''+2b)}\left[\frac{\phi'^2}{\phi^2}\right]''+\frac{a_2}{(3b''+2b)}\left(\frac{a''}{a}-\frac{a'^2}{a^2}\right)\frac{\phi'^2}{\phi^2}+\frac{2a_2}{(3b''+2b)}\frac{a'}{a}\frac{\phi'\phi''}{\phi^2}-\nonumber\\
& &-\frac{2a_2}{(3b''+2b)}\frac{a'}{a}\frac{\phi'^3}{\phi^3}=0,
\nonumber \\
              &   & \nonumber \\
              &   &C\left[ \ln a\; \phi'''' + (\ln a\; \phi)''''\right]-4a_1\frac{a'}{a}\frac{\phi'^3}{\phi^4}+2a_2\frac{a''}{a}\left(\frac{\phi'^2}{\phi^3}-\frac{\phi''}{\phi^2}\right)+\nonumber \\ & &+a_2\left(2\frac{a'a''}{a^2}-\frac{a'''}{a}\frac{\phi'}{\phi^2}\right)- 12a_1\left[\frac{\phi'^2\phi''}{\phi^4}-\frac{\phi'^4}{\phi^5}\right]\ln a=0,
\end{eqnarray}
 where $3 b''+2 b \neq 0$. The natural choice for $b''$ is to take $b''=0$, since the
choice of $b''$ doesn't make any difference in the physical
effects.

In order to solve these equations of motion let us make the
transformation to cosmological time in the above equations
\cite{brevik99} $ dt = a(\eta)\,d\eta $. Then the first of the
equations of motion (\ref{eqmov}) takes the following form:
\begin{eqnarray}
\!\!\!\!\!\!\!\!\!\!& & \!\!\!\!\!\!\!\!\!\!a^2 \stackrel{....}{a}
+ 3a \dot{a}\stackrel{...}{a}+a \ddot{a}^2-\left( 5+\frac{6 b'}{b}
\right) \dot{a}^2\,\ddot{a} -\frac{3}{\kappa b}(a^2 \ddot{a}+a
\dot{a}^2) + \frac{6}{\kappa b}a^3\Lambda - \frac{C \phi
Y[\phi,a]}{4b}-\nonumber \\ &
 &\!\!\!\!\!-\frac{a_1}{4b}\frac{a^3\dot{\phi}^4}{\phi^4}-2a_2a^2\left[a\frac{\ddot{\phi}^2}{\phi^2}+3\dot{a}\frac{\dot{\phi}\ddot{\phi}}{\phi^2}+a\frac{\dot{\phi}\stackrel{...}{\phi}}{\phi^2}
-5a\frac{\dot{\phi}^2\ddot{\phi}}{\phi^3}-3\dot{a}\frac{\dot{\phi}^3}{\phi^3}+3a\frac{\dot{\phi}^4}{\phi^4}\right]=0.\qquad
\label{9}\end{eqnarray}

Here,
\begin{eqnarray*}
Y[\phi,a]&= & a^3 \stackrel{....}{\phi}+6a^2 \dot{a}\stackrel{...}{\phi} \\
         &+ & 4a^2\ddot{a}\ddot{\phi}+7a \dot{a}^2\ddot{\phi}+4a \dot{a}\ddot{a}\dot{\phi}
+a^2\stackrel{...}{a}\dot{\phi} +\dot{a}^3 \dot{\phi}.
\end{eqnarray*}
The second of Eqs.(\ref{eqmov}) becomes:
\begin{eqnarray}
\!\!\!\!\!\!\!\!\!\!& & \!\!\!\!\!\!\!\!\!\!C \{2 a \ln a
\;Y[\phi,a]+\phi
a^3\stackrel{....}{a}+4a^3\dot{a}\stackrel{...}{\phi}
+3a^2\phi\,\dot{a}\stackrel{...}{a}+4a^3\dot{\phi}
\stackrel{...}{a}+6a^3\ddot{\phi}\ddot{a}+12a^2\ddot{\phi}\;\dot{a}^2 +\nonumber \\
\!\!\!\!& & \!\!\!\!
 +14a^2\dot{\phi}\dot{a}\ddot{a}+a\phi\ddot{a}\dot{a}^2+a^2\phi\ddot{a}^2
+4a\dot{\phi}\dot{a}^3 \} + \nonumber \\\!\!\!\!\!& & \!\!\!\!\!
+12a_1\left[a\frac{\dot{\phi}^4}{\phi^5}-a\frac{\dot{\phi}^2\ddot{\phi}}{\phi^4}-\frac{\dot{\phi}^3}{\phi^4}\right]a^3\ln
a - 4a_1a^3\dot{a}\frac{\dot{\phi}^3}{\phi^4}+ \nonumber \\
\!\!\!\!& & \!\!\! + a_2
a\big[2a^2\ddot{a}\frac{\dot{\phi}^2}{\phi^3}-2a^2\ddot{a}\frac{\ddot{\phi}}{\phi^2}+2a\dot{a}^2\frac{\dot{\phi}^2}{\phi^3}-2a\dot{a}^2\frac{\ddot{\phi}}{\phi^2}
- 2a^2\stackrel{...}{a}\frac{\dot{\phi}}{\phi^2} -\nonumber \\
\!\!\!\!& & \!\!\!- 8a\dot{a}\ddot{a}\frac{\dot{\phi}}{\phi^2}-
\dot{a}^3\frac{\dot{\phi}}{\phi^2}\big]=0.
\label{eqmov1}\end{eqnarray}

These equations are too complicated to be solved analytically in
general. To construct the solution for these equations we will
consider only the case with the presence of the dilaton. Other
cases which have been considered in \cite{brevik99, brevik,
quiroga} will not give us any new information. It is important
only to remark that if we consider the case without dilaton, we
will obtain the same solution which leads to the inflationary
solution (for positive $H$) obtained first by Starobinsky in
\cite{Sta} using the renormalized EMT of conformal matter on the
right-hand side of Einstein's equations.

Now let us consider the most interesting case, when we have a
non-zero dilaton. An approximated special solution of Eqs.
(\ref{eqmov}) when the term with $\ln a$ may be dropped, may be
obtained. As well $a(t)=a_0e^{Ht}$, $\ln a \sim Ht$.Moreover, $H$
is proportional to the Planck mass, so $Ht$ is small quantity so
our approximation is justified.

Then, we search for special solutions of the sort:
\begin{equation}
a(t) \simeq \tilde{a}_0 e^{\tilde{H}t},~~~~\phi(t) \simeq \phi_0
e^{-\alpha \tilde{H}t}
\label{11}\end{equation}

As we said before, since we know that that both $N$ and $M$ maybe
considered to be big parameters, it means  one can study large-$N$
or large-$M$ expansion. So we will consider two cases as follows:

\begin{itemize}
\item Case $M\gg N^2$, when one may neglect in $b',\quad b$ all
the terms with $N$ and so term with $C$. This case coincides with
the one considered in \cite{geyer}, in that work the main
difference is that the classical action includes the kinetic terms
on the dilaton, and in our theory this term is absent. If we
included it we would find a different kinematics for our model.

According to this and substituting these solutions (\ref{11}) in
Eq.(\ref{eqmov}) and neglecting the logarithmic term and term with
parameter $C$, we obtain for $\alpha$ the following values,

\begin{eqnarray}
\alpha_0 = 0, \qquad \alpha_1= \frac{3\sqrt{3}}{4}i, \qquad
\alpha_2= -\frac{3\sqrt{3}}{4}i \label{bigM}\,.\end{eqnarray}

From these solutions we find for the dilaton the following
solution,

\begin{eqnarray}
\phi=\phi_0+\phi_1\sin{\alpha Ht}+\phi_2\cos{\alpha Ht}
\label{phim}\,.\end{eqnarray}

This solution oscillates and moreover if we take $\phi_0, \quad
\phi_1$ and $\phi_2$ to be small, the oscillation will be about
zero.

From another side the solution for $H^2$ is

\begin{eqnarray}
\label{13M} \tilde{H}^2 \simeq -\frac{3}{2kb'}\frac{1+
\sqrt{1+\frac{8k\Lambda}{3}\left[\Big(
6b'+\frac{a_1\alpha^4}{4}\Big)\right]}}{6+\frac{a_1\alpha^4}{4b'}
}\,.
\end{eqnarray}

Hence we find that although the solution for the dilaton
oscillates, it is possible to obtain a non imaginary value for the
scalar factor, which describes dilatonic de Sitter Universe.

\item Case $N^2\approx M$.

Substituting solutions (\ref{11}) in Eq.(\ref{eqmov}) and
neglecting the logarithmic term, we obtain for $\alpha$ the
following equation,

\begin{eqnarray}
\alpha^2+\Big(\frac{1}{48\phi_0^2}-\frac{11}{9}\Big)\alpha +
\frac{1}{3} = 0 \label{12}\,.\end{eqnarray}

After solving this equation (\ref{12}) by taking for the explicit
example, values $\phi_0< 0.093693$ or $\phi_0> 0.558694$, $\alpha$
will be always real quantity and therefore the solution for the
dilaton does not oscillate. In other cases we will have an
oscillating dilaton.

For $H^2$ , by performing similar analysis like before,it is not
difficult to obtain the following solution,

\begin{eqnarray}
\label{13} \tilde{H}^2 \simeq -\frac{3}{2kb'}\frac{\!\!1\!\!+
\sqrt{1+\frac{8k\Lambda}{3}\left[6b'+\frac{a_1\alpha^4}{4}+\frac{\big(\alpha^4-6\alpha^3+11\alpha^2-6\alpha
\big)C\phi_0^2}{4} \right]}}{6+\frac{a_1\alpha^4}{4b'}
+\frac{C\phi_0^2}{4b'}\big( \alpha^4-6\alpha^3+11\alpha^2-6\alpha
\big)}\,.
\end{eqnarray}

 This solution, for any choice of $b''$, and choosing the values for $N$ , $M$ and $\phi_0$   is always
 positive.This shows that there are many possibilities to choose
 the parameters so that dilatonic de Sitter quantum universe
occurs.
\end{itemize}

It is easy to verify that obtained solution, in both cases,
(\ref{13}) and (\ref{13M}) is bigger than the obtained in the case
of no dilaton (see \cite{brevik99}). So we conclude , that in this
case, the role of the dilaton is to make the inflation faster, as
compared with the case of no dilaton.

Let us now discuss how the stability of the inflation depends on
the pertubation for the scale factor $a$. In order to answer to
this question we look for a solution for $a$ like $a = \tilde{a}_0
e^{\tilde{H}t}+ a_1$. Substituting this solution into the
equations of motion (\ref{eqmov}), we obtain the following
equation for the perturbation theory.

\begin{equation}
a_1'' + (H^2-2\Lambda)a_1=0 \label{eqper}\end{equation}

As is seen the solutions for this equation depends on the sign of
its second term, so we have,

\begin{eqnarray}
\label{osc} a_1=R\sin \left[ (H^2-2\Lambda)t+r_0 \right] & if &
H^2-2\Lambda > 0\\
\label{exp} a_1=\frac{a_0}{2}\left[ e^{(H^2-2\Lambda)t}+
e^{-(H^2-2\Lambda)t} \right] & if & H^2-2\Lambda <
0\,.\end{eqnarray}

Here $R, r_0$ in (\ref{osc}) are some constants, which depend on
the initial conditions. For the second solution (\ref{exp}) the
initial conditions $a_1(0)=a_0, \dot a_1(o)=0$ have been used.
From these solutions it is easy to conclude that the inflation is
stable since in the first case we have that it oscillates with a
very small frequency of oscillation around  the value of $a_0$ and
in the second case the inflation grows as an exponent with a very
small speed, from the value of $a_0$. In other words, such
Universe could be the eternal dilatonic de Sitter Universe.

    Summarizing , we have discussed quantum cosmology from $\cal{N}
    $ = 4 super YM theory on a dilaton-gravitational background where super symmetry is broken by scalar fields,
    taking into account, in the classical action, the cosmological constant.
    As one solution we found the inflationary (conformally flat)
    Universe with exponentially decaying dilaton. It is not
    difficult to consider other cosmogical models (say closed or
    open FRW), or other types of behaviour of the scale factor.
    However in this case one should do a numerical study of the
    effective equations of motion. The results, of course, should
    depend very much on the choice of initial conditions for
    $a(t), ~\phi(t)$ and their derivatives. We have also seen that
    for our theory, the inflation has a very stable behaviour.

The very interesting generalization of this work is related with
situation where $4d$ dilaton coupled spinors are included. (For
recent study of spinors from group theoretical point of view, see
\cite{ahluwalia}). The corresponding conformal anomaly is found in
\cite{PNSN}. Then, one can consider supersymmetric system
consisting of $N=4$ SYM and dilaton-coupled Wess - Zumino model
\cite{SJG}. Cosmology of such a model will be described elsewhere.

Finally, let us note that trace anomaly driven cosmology as above can be easily related with quantum brane cosmology (for recent review, see \cite{sdo})\\

{\bf Acknowledgments} I am grateful to Prof. S.D. Odintsov for
formulation of the problem and numerous helpful discussions. I
thank Prof. P.M. Lavrov for very useful discussions. The research
of J. Q. H.(professor at Unviversidad Tecnol\'ogica de Pereira),
has been supported by Professorship and Fellowship from the
Universidad Tecnol\'ogica de Pereira, Colombia.

\end{document}